\documentclass[11pt]{article}
\usepackage{fullpage,epsf,amssymb,amsthm,amsfonts,amsmath,latexsym,array,graphicx,extarrows,mathtools,mathstyle,mathrsfs,slashed,subcaption,amsbsy,csquotes,accents}
\usepackage[normalem]{ulem}
\usepackage{authblk}
\usepackage{cite}
\usepackage{color}    
\usepackage{float}
\usepackage{hyperref}
\usepackage[export]{adjustbox}

\numberwithin{equation}{section}

\title{\bf{Pion spectral properties above the chiral crossover \\ of QCD}} 

\author{Peter Lowdon}
\author{Owe Philipsen}

\affil{{\scriptsize Institut f\"{u}r Theoretische Physik, Goethe-Universit\"{a}t, Max-von-Laue-Str. 1,  60438 Frankfurt am Main, Germany}}

\date{}
\begin{document}
\maketitle

\begin{abstract}
\noindent
Spectral functions encode a wealth of information about the dynamics of any given system, and the determination of their non-perturbative characteristics is a long-standing problem in quantum field theory. Whilst numerical simulations of lattice QCD provide ample data for various Euclidean correlation functions, the inversion required to extract spectral functions is an ill-posed problem. In this work, we pursue previously established constraints imposed by field locality at finite temperature $T$, namely that spectral functions possess a non-perturbative representation which generalises the well-known K\"{a}ll\'{e}n-Lehmann spectral form to $T>0$. Using this representation, we analyse lattice QCD data of the spatial pseudo-scalar correlator in the temperature range $220-960 \, \text{MeV}$, and obtain an analytic expression for the corresponding spectral function, with parameters fixed by the data. From the structure of this spectral function we find evidence for the existence of a distinct pion state above the chiral pseudo-critical temperature $T_{\text{pc}}$, and contributions from its first excitation, which gradually melt as the temperature increases. As a non-trivial test, we find that the extracted spectral function reproduces the corresponding temporal lattice correlator data for $T=220 \, \text{MeV}$.
\end{abstract}

\newpage

\section{Introduction}
\label{intro}

The phases of QCD under extreme conditions and the nature of its associated effective degrees of freedom are among the most pressing problems of theoretical physics, affecting experimental programs from heavy ion collisions, to astro-particle and gravitational wave physics. Of particular interest is the question, deeply related to the confinement problem, of how ordinary hadronic matter gets modified in medium to eventually dissolve into the expected quark gluon plasma. In principle, the answer is provided by the spectral properties of Euclidean two-point functions of gauge-invariant operators $O_\Gamma(\tau,\vec{x})$
\begin{align}
C_{\Gamma}(\tau,\vec{x}) = \langle O_{\Gamma}(\tau,\vec{x})\,O_{\Gamma}(0,\vec{0})\rangle_{T},
\label{eq:corr0}
\end{align}
where $\Gamma$ denotes a set of quantum numbers, and the expectation value is over a thermal ensemble at temperature $T=1/\beta$, which for the purposes of this work we specialise to zero baryon density, $\mu_B=0$. The spatial Fourier transform of the correlation functions take the universal form 
\begin{align}
\widetilde{C}_\Gamma(\tau,\vec{p}) = \int_{0}^{\infty} \frac{d\omega}{2\pi} \frac{\cosh\left[\left(\tfrac{\beta}{2}-|\tau| \right)\omega\right] }{\sinh\left(\tfrac{\beta}{2}\omega\right)} \,\rho_{\Gamma}(\omega,\vec{p}), 
\label{eq:corr}
\end{align}
where the associated spectral functions $\rho_{\Gamma}(\omega,\vec{p})$ contain the desired information about the possible excitations in a given quantum number channel. \\

\noindent
In order to fully describe a thermal system of strong-interaction particles one ultimately requires a non-perturbative approach. Lattice QCD is a powerful such approach, and numerical computations of the correlation functions in Eq.~\eqref{eq:corr0} have led to significant advancement in the understanding of finite-temperature phenomena~\cite{Detar:1987kae,Detar:1987hib,Born:1991zz,Florkowski:1993bq,Kogut:1998rh,Aarts:2005hg,Wetzorke:2001dk,Karsch:2003jg,Petreczky:2003iz,Asakawa:2003re,Aarts:2006em, Ding:2012sp,Burnier:2015tda,Mukherjee:2015mxc,Meyer:2017ydp,Rothkopf:2019ipj,Spriggs:2021dsb}. On the lattice it is the \textit{spatial} correlators, integrated over the orthogonal directions, that are accessible over the largest distances and hence contain the most information about the thermal system. Particular progress has been made in establishing the properties of these correlators in recent years~\cite{Laermann:2001vg,Wissel:2005pb,Cheng:2010fe,Banerjee:2011yd,Karsch:2012na,Brandt:2014uda,Bazavov:2014cta,DallaBrida:2021ddx}, which are defined as  
\begin{align}
C_{\Gamma}(x_{3}) &= \int^{\infty}_{-\infty} \! dx_{1} \int^{\infty}_{-\infty} \! dx_{2} \int_{-\frac{\beta}{2}}^{\frac{\beta}{2}} d\tau \, C_{\Gamma}(\tau, \vec{x}) \label{C_def} \\
&= \int_{-\infty}^{\infty}  \frac{dp_{3}}{2\pi}e^{i p_{3} x_{3}} \int_{0}^{\infty}  \frac{d\omega}{\pi \omega}  \ \rho_{\Gamma}(\omega,p_{1}=p_{2}=0,p_{3}),  \label{C_rho}
\end{align}
where the second equality follows from Eq.~\eqref{eq:corr}. Equation~\eqref{C_rho} demonstrates that the structure of $C_{\Gamma}(x_{3})$ is entirely controlled by the spectral function of the corresponding hadronic operator, and thus directly probes the spectral properties of QCD for $T>0$. A fundamental issue with Eq.~\eqref{eq:corr} though is that the extraction of the corresponding spectral function is an ill-posed inverse problem. For this reason, most strategies to obtain QCD spectral functions from lattice data require intricate statistical methods, combined with input based on either perturbative calculations or phenomenological modelling. A general discussion of such strategies can be found in Refs.~\cite{Asakawa:2000tr,Meyer:2011gj}. \\

\noindent
In this work we pursue a different approach, which was developed in Refs.~\cite{Bros:1992ey,Bros:1995he,Bros:1998ua,Bros:1996mw,Bros:2001zs} and is based on the $T>0$ generalisation of axiomatic formulations of local vacuum-state QFT, whose applications over the years have led to numerous foundational insights, including the relationship between spin and statistics, the generality of \textit{CPT} symmetry, and the rigorous connection of Minkowski and Euclidean QFTs~\cite{Streater:1989vi,Haag:1992hx,Bogolyubov:1990kw}. The non-perturbative $T>0$ framework of Refs.~\cite{Bros:1992ey,Bros:1995he,Bros:1998ua,Bros:1996mw,Bros:2001zs} focussed on the simplest case of Hermitian scalar fields $\phi(x)$, and established that characteristic features such as the loss of Lorentz symmetry can be incorporated by defining a thermal background state $|\Omega_{\beta}\rangle$ at temperature $T=1/\beta$, which is no longer invariant under the full Poincar\'{e} group. Together with the standard constraints brought about by the assumption of thermal equilibrium~\cite{Kapusta:2006pm,Bellac:2011kqa}, it was demonstrated that the \textit{locality} of the fields\footnote{By locality we mean: $\left[\phi(x),\phi(y)\right]=0$ for $(x-y)^{2}<0$, which is simply the physical assumption that all measurements respect causality.} alone imposes particularly significant constraints, and implies the following representation of the scalar spectral function\footnote{As is standard in the literature, the spectral function refers to the Fourier transform of the thermal two-point commutator $\langle \Omega_{\beta}| \left[\phi(x),\phi(y)\right]|\Omega_{\beta} \rangle$.}~\cite{Bros:1992ey}
\begin{align}
\rho(\omega,\vec{p}) = \int_{0}^{\infty} \! ds \int \! \frac{d^{3}\vec{u}}{(2\pi)^{2}} \ \epsilon(\omega) \, \delta\!\left(\omega^{2} - (\vec{p}-\vec{u})^{2} - s \right)\widetilde{D}_{\beta}(\vec{u},s).
\label{commutator_rep}
\end{align}
In the $T\rightarrow 0$ limit Eq.~\eqref{commutator_rep} reduces to the well-known K\"{a}ll\'{e}n-Lehmann spectral representation\footnote{For the vacuum spectral function the K\"{a}ll\'{e}n-Lehmann representation has the momentum-space form~\cite{Kallen:1952zz,Lehmann:1954xi}: $2\pi \epsilon(\omega)  \int_{0}^{\infty} \! ds \, \delta\!\left(p^{2} - s \right) \rho(s)$, where $\rho(s)$ is the spectral density whose singularities capture the presence of stable particle states.}, and hence represents its $T>0$ generalisation. From the structure of Eq.~\eqref{commutator_rep} one can see that the effects of the background state are entirely captured by the \textit{thermal spectral density} $\widetilde{D}_{\beta}(\vec{u},s)$. Determining the properties of this quantity is therefore essential for describing the characteristics of scalar particles in thermal media. \\ 

\noindent
Equation~\eqref{commutator_rep} is completely general and holds for any scalar field satisfying locality. In order to understand the characteristics of the thermal spectral density for specific theories, additional information is necessary. In Ref.~\cite{Bros:1992ey} the authors proposed that the singular structure of $\widetilde{D}_{\beta}(\vec{u},s)$ in the variable $s$ is preserved relative to the vacuum theory, as long as no phase transition is met. This implies that discrete and continuous contributions are decomposed, and hence if a theory contains a stable particle state of mass $m$ at $T=0$ one can write~\cite{Bros:2001zs}
\begin{align}
\widetilde{D}_{\beta}(\vec{u},s)= \widetilde{D}_{m,\beta}(\vec{u})\, \delta(s-m^{2}) + \widetilde{D}_{c, \beta}(\vec{u},s),
\label{decomp}
\end{align} 
where $\widetilde{D}_{c, \beta}(\vec{u},s)$ is continuous in $s$. As outlined in Refs.~\cite{Bros:1992ey,Bros:1995he,Bros:1998ua,Bros:1996mw,Bros:2001zs}, there are several reasons for why the discrete component in Eq.~\eqref{decomp} provides a natural description of a particle state in finite-temperature QFT. Firstly, for $T>0$ the so-called \textit{damping factor} $\widetilde{D}_{m,\beta}(\vec{u})$ is non-trivial, which due to the structure of Eq.~\eqref{commutator_rep} causes $\rho(\omega,\vec{p})$ to have contributions outside of the mass shell $p^{2}=m^{2}$, and hence the $T=0$ mass peak of the particle becomes broadened, as one would expect. Moreover, the precise nature of this broadening has been shown to be controlled by the underlying interactions between the particle state and the constituents of the thermal medium~\cite{Bros:2001zs}. Although Eq.~\eqref{decomp} assumes the $T=0$ particle state to be stable, this can in principle be generalised to unstable states by replacing $\delta(s-m^{2})$ with a suitable resonance-type function in $s$, such as a relativistic Breit-Wigner. The factorisation of the $(\vec{u},T)$ and $s$ dependence therefore ensures that this representation can distinguish between particle decays of different physical origin, namely those brought about by dissipative thermal effects, controlled by $(\vec{u},T)$, and those due to any intrinsic instability of the $T=0$ particle states. The structure of damping factors in specific models was first explored in Ref.~\cite{Bros:2001zs}, and more recently in Refs.~\cite{Lowdon:2021ehf,Lowdon:2022keu,Lowdon:2022ird}, where it was shown that these quantities can in fact be used to perform non-perturbative analytic calculations of in-medium observables, including the shear viscosity. \\

\noindent
The goal of the present work is to apply the observations from Refs.~\cite{Lowdon:2022keu,Lowdon:2022ird} to lattice data for pseudo-scalar correlators in order to compute the thermal spectral density $\widetilde{D}_{\beta}(\vec{u},s)$, and from it the spectral function $\rho(\omega,\vec{p})$. In Sec.~\ref{extract} we establish an analytic connection between the spatial correlator and the spectral representation in Eq.~\eqref{commutator_rep}. In Sec.~\ref{scalar_QCD} we use these results to extract the properties of the light-quark pseudo-scalar spectral function from lattice QCD data, and discuss the physical implications of these results. Finally, in Sec.~\ref{concl} we summarise our main findings.

\section{Spectral representation of spatial correlators}
\label{extract}

We begin by discussing the relation between the thermal spectral density $\widetilde{D}_{\beta}(\vec{u},s)$ and the imaginary-time two-point function $C(\tau, \vec{x})$. In Ref.~\cite{Lowdon:2022keu} it was shown for scalar fields that this connection is defined via the following integral relation  
\begin{align}
\int_{-\frac{\beta}{2}}^{\frac{\beta}{2}} d\tau \,C(\tau, \vec{x}) =\frac{1}{4\pi |\vec{x}|}\int_{0}^{\infty} \! ds \ e^{-|\vec{x}|\sqrt{s}} D_{\beta}(\vec{x},s),
\label{W_int}
\end{align}
where $D_{\beta}(\vec{x},s)$ is the inverse Fourier transform of $\widetilde{D}_{\beta}(\vec{u},s)$. Using the fact that $D_{\beta}(\vec{x},s)$ depends only on $|\vec{x}|$ in an isotropic medium, it follows by combining Eqs.~\eqref{C_def} and~\eqref{W_int} that the spatial correlator has the general spectral representation
\begin{align}
C(x_{3}) = \frac{1}{2}\int_{0}^{\infty} \! ds \int^{\infty}_{|x_{3}|} \! dR \ e^{-R\sqrt{s}} D_{\beta}(R,s).
\label{C_int}
\end{align}
\ \\
\noindent
As outlined in Sec.~\ref{intro}, in the simplest case where there exists a single stable particle state at $T=0$, there is evidence to suggest~\cite{Bros:2001zs} that the thermal spectral density has the decomposition in Eq.~\eqref{decomp}. However, hadronic correlators in QCD in general receive contributions from multiple states. In this case, the decomposition in Eq.~\eqref{decomp} has the natural position-space generalisation 
\begin{align}
D_{\beta}(\vec{x},s)= \sum_{i} D_{m_{i},\beta}(\vec{x})\, \delta(s-m_{i}^{2}) + D_{c, \beta}(\vec{x},s),
\label{D_decomp}
\end{align}
where $m_{i}$ are the $T=0$ masses associated with these states. Using analogous arguments to those in Ref.~\cite{Lowdon:2022keu}, if $D_{c, \beta}(\vec{x},s)$ is non-vanishing for $s \geq s_{c}$, and the gap between the scale $s_{c}$ and some subset of the masses $\{m_{1},..., m_{n}\}$ is sufficiently large, the damping factors associated with these states will dominate the behaviour of $C(x_{3})$
\begin{align}
C(x_{3}) \approx \frac{1}{2}\sum_{i=1}^{n} \,  \int^{\infty}_{|x_{3}|} \! dR \  e^{-m_{i}R} D_{m_{i},\beta}(R),
\label{C_decomp_dom}
\end{align}
and this domination will be especially pronounced at large $x_{3}$. Equation~\eqref{C_decomp_dom} demonstrates not only that 
the vacuum particle states have a significant impact on the behaviour of $C(x_{3})$, but also how this behaviour is connected to the dynamical properties of the theory. In the situation that one of the masses $m_{1}$ is significantly smaller than the others, it follows from Eq.~\eqref{C_decomp_dom} that the corresponding damping factor $D_{m_{1},\beta}$ associated with this state can be directly calculated from the leading large-$x_{3}$ behaviour of the spatial correlator derivative
\begin{align}
D_{m_{1},\beta}(|\vec{x}|=x_{3}) \sim -2e^{m_{1} x_{3}} \, \frac{d C(x_{3})}{dx_{3}}, \quad\quad x_{3} \rightarrow \infty. 
\label{D_C_rel}
\end{align}
In principle, the damping factors associated with the higher mass states $\{m_{2},..., m_{n}\}$ can also be extracted, in this case from the sub-leading large-$x_{3}$ behaviour of $C(x_{3})$. The practical feasibility of this depends on the relative size of the masses and the behaviour of $D_{m_{i},\beta}(\vec{x})$. Once explicit particle damping factors are available, one can then use Eq.~\eqref{commutator_rep} to calculate their analytic contributions to the full spectral function $\rho(\omega,\vec{p})$.
 
\section{Scalar QCD correlators}
\label{scalar_QCD}

\subsection{Preliminary considerations}
\label{prelim}

Before applying the relations derived in the last section to lattice QCD data, let us recall a result from the literature that explicitly supports the validity of the ansatz in Eq.~\eqref{decomp} in QCD, at least for low temperatures and in the chiral limit. In Refs.~\cite{Dey:1990ba,Eletsky:1992ay} QCD vector and axial-vector correlators were evaluated using the well-tested PCAC current algebra, which to leading order in $\epsilon=T^2/(6f_\pi^2)$ allows one to express the low-temperature correlators by the vacuum correlators   
\begin{align}
C_{V}(p,T) &= (1-\epsilon)C_{V}(p,0)+\epsilon \, C_{A}(p,0), \nonumber  \\
C_{A}(p,T) &= (1-\epsilon)C_{A}(p,0)+\epsilon \, C_{V}(p,0).
\label{ioffe}
\end{align}
Whilst these are not scalar correlators, they provide particular examples for the more general assumption leading to Eq.~\eqref{decomp}, namely that certain analytic structures of vacuum correlators are preserved at finite temperatures. In the case of Eq.~\eqref{ioffe}, the poles of the vacuum correlators are mixed, but explicitly present in the finite-temperature correlators. Similar considerations apply, in less detail but more generality, to the spectral decomposition of lattice meson correlators, 
\begin{align}
C_\Gamma(\tau,\vec{x}) = \frac{1}{\mathcal{Z}}\sum_{m,n}|\langle m| O_\Gamma(0,\vec{x}) |n\rangle|^2 \; e^{-\tau E_{n}}e^{-(\beta-\tau)E_{m}},
\end{align}
where $\mathcal{Z}$ is the partition function, and $H$ has the discrete spectrum $H|n\rangle = E_n|n\rangle$. Since the Hamiltonian is independent of $T$, this spectrum is identical to that in the vacuum, as long as no non-analytic phase transition is crossed. Again, it is evident that the finite-temperature correlator inherits the vacuum particle states and their analytic structures, which get mixed with weight factors determined by temperature\footnote{Incidentally, the hadron resonance gas model, which is known to provide a reasonable description of the lattice QCD equation of state up to the chiral crossover at $T_{pc}\approx 155$ MeV, employs exclusively masses and decay widths encoded in vacuum correlation functions.}. Since there is no non-analytic phase transition in QCD at $\mu_{B}=0$, the proposition in Eq.~\eqref{decomp} appears well suited for lattice pion correlators over a large temperature range, and we shall now use it to extract the corresponding spectral information.

\newpage
 
\subsection{Pseudo-scalar lattice correlator analysis}
\label{lattice}

\begin{figure}[t]
\centering
\includegraphics[width=0.65\columnwidth]{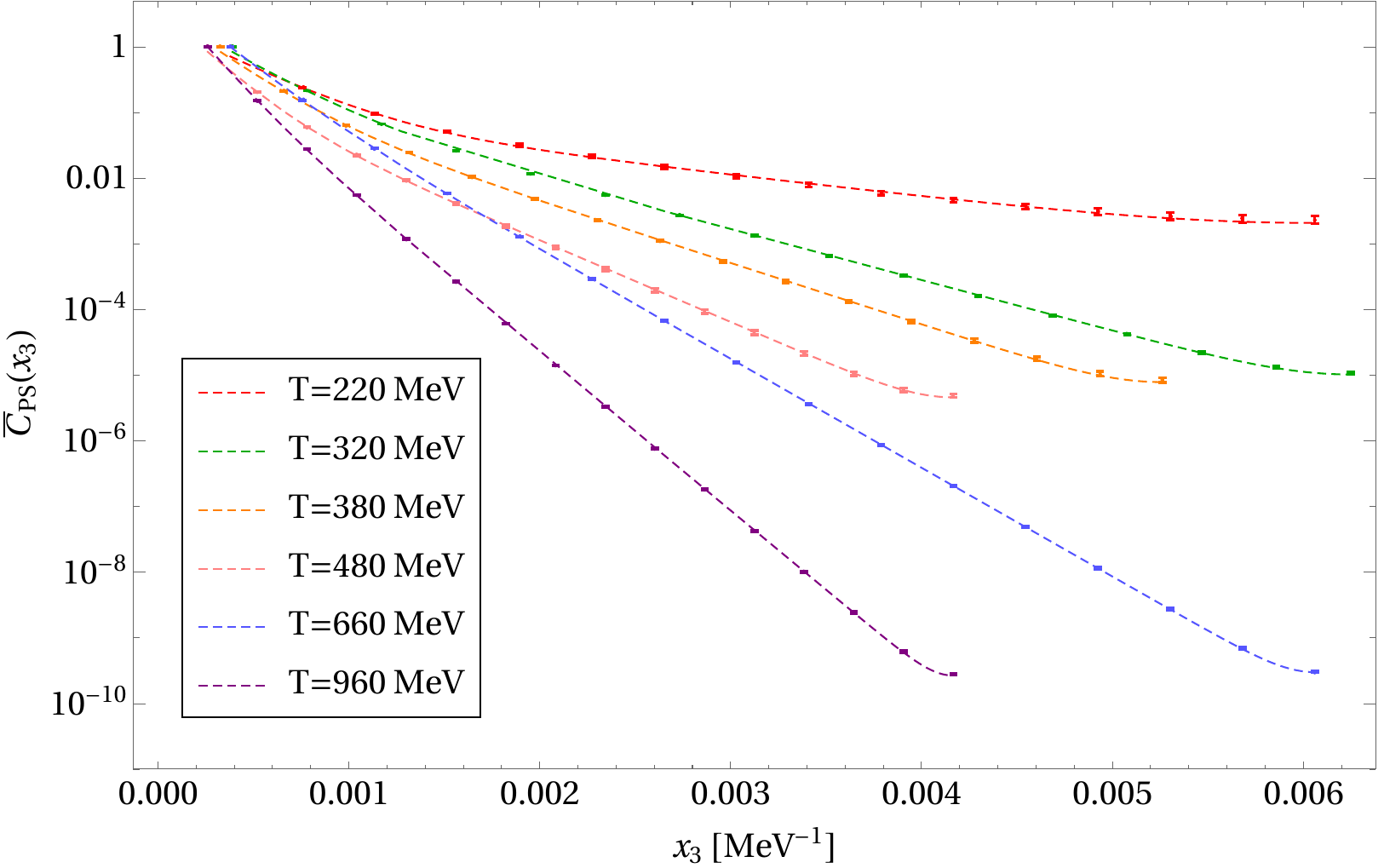}
\caption{Spatial pseudo-scalar lattice correlators obtained with $N_{f}=2$ domain wall fermions in \cite{Rohrhofer:2019qwq}. The dashed lines represent fits to the data using the ansatz in Eq.~\eqref{2-state}. The corresponding $\chi^{2}/\text{d.o.f.}$ of the fits are listed in Table~\ref{tab:ensembles}.} 
\label{z_corr_plot} 
\end{figure} 
 
We now focus on spatial pseudo-scalar correlators defined as
\begin{align}
C_{\text{PS}}^{a}(x_{3}) =  \int^{\infty}_{-\infty} \! dx_{1} \int^{\infty}_{-\infty} \! dx_{2} \int_{-\frac{\beta}{2}}^{\frac{\beta}{2}} d\tau \, \langle \Omega_{\beta}|O_{\text{PS}}^{a}(\tau, \vec{x})O_{\text{PS}}^{a \, \dagger}(0)|\Omega_{\beta}\rangle,
\label{PS_C_def}
\end{align} 
where $O_{\text{PS}}^{a}= \overline{\psi}\gamma_{5}\frac{\tau^{a}}{2}\psi$, with $\psi(x)=(u(x),d(x))$ isospin doublets, and $\tau^{a}$ the Pauli spin matrices. For the remainder of the paper the isospin index $a$ will be dropped since the spatial correlator is independent of $a$ for degenerate quark masses. We analyse the lattice data generated in Ref.~\cite{Rohrhofer:2019qwq} with two mass-degenerate flavours of domain wall fermions, which have good chiral symmetry properties also at finite lattice spacing. Thermal ensembles for $T=$ 220, 320, 380, 480, 660, and 960~MeV were generated with a fixed bare quark mass of $am_{u}=am_{d}=0.001$, which is compatible with the physical pion mass at $T=0$. The correlator in Eq.~\eqref{PS_C_def} was evaluated in the range $x_{3}\in[x_{3}^{*},L/2]$, where $L$ is the spatial lattice extent, and $x_{3}^{*}=1/ (T N_{\tau})$ is the smallest non-vanishing separation by one lattice spacing, with $N_{\tau}$ the temporal lattice extent. The calculation in Ref.~\cite{Rohrhofer:2019qwq} was also performed such that the correlator was normalised to one at $x_{3}=x_{3}^{*}$. To avoid ambiguity we will refer to this normalised correlator as $\overline{C}_{\text{PS}}(x_{3})$. More details regarding the precise lattice setup can be found in Ref.~\cite{Rohrhofer:2019qwq} and a brief summary is given in Table~\ref{tab:ensembles}. \\

\noindent
We begin our analysis by parametrising the functional form of the correlator $\overline{C}_{\text{PS}}(x_{3})$. In order to do so, one must first understand the particle states that are generated by the light quark pseudo-scalar operator $O_{\text{PS}}^{a}$. For masses below 2~GeV, these states consist of the pion $\pi$, $\pi(1300)$, and $\pi(1800)$~\cite{Workman:2022ynf}, the latter two of which are generally interpreted as being the first two radial excitations of the pion, the $\pi^{*}$ and $\pi^{**}$~\cite{Holl:2004fr}. In accordance with numerous previous investigations, we find the large $x_{3}$ behaviour of $\overline{C}_{\text{PS}}(x_{3})$ to be consistent with a pure exponential. Taking into account the periodic boundary conditions along the $x_{3}$ direction, we thus started by fitting the standard functional form 
\begin{align}
\overline{C}_{\text{PS}}(x_{3}) = A e^{-m_{\pi}^{\text{scr}} x_{3}}+ A e^{-m_{\pi}^{\text{scr}} (L-x_{3})}
\end{align}
to a range $x_{3}\in[x_{3}^{min},L/2]$, where $m_{\pi}^{\text{scr}}$ defines the so-called screening mass of the lowest energy state, in this case the pion. The lower boundary of the fit range $x_{3}^{min}$ was chosen such that the inclusion of an additional data point would significantly reduce the quality of the fit, producing an $\mathcal{O}(1)$ increase in the $\chi^{2}/\text{d.o.f.}$ By adding an additional exponential component to our fit ansatz,
\begin{align}
\overline{C}_{\text{PS}}(x_{3}) = A e^{-m_{\pi}^{\text{scr}} x_{3}}+ A e^{-m_{\pi}^{\text{scr}} (L-x_{3})} + B e^{-m_{\pi^{*}}^{\text{scr}} x_{3}}+ B e^{-m_{\pi^{*}}^{\text{scr}} (L-x_{3})},
\label{2-state}
\end{align}
the overall quality of the fit was significantly improved, enabling excellent fits over the entire data range $[x_{3}^{*},L/2]$ for $T=$ 660 MeV and 960~MeV, and excluding only
the smallest separation point for $T=$ 220, 320, 380, and 480~MeV. These data and the fits are shown in Fig.~\ref{z_corr_plot}. The corresponding fit parameters, together with their uncertainties, are listed in Table~\ref{tab:ensembles}. \\

\noindent
In Eq.~\eqref{2-state} the exponents of the second pair of terms have the interpretation of a screening mass for the second-lightest pseudo-scalar meson state, the $\pi^{*}$. As opposed to the pion, the $\pi^{*}$ is unstable under strong interactions, but this is not observed in the data, presumably due to an insufficient lattice resolution. The exact correlator can also contain additional contributions from higher excited states, including the $\pi^{**}$, although these states are increasingly exponentially suppressed. One could in principle extract the properties of these states from the short distance behaviour of the correlator on finer lattices with higher resolution and more lattice data points, but at the current resolution these states do not play any role. As an independent check, the resulting pion screening masses (listed in Table~\ref{tab:ensembles}) can be compared with the continuum extrapolated values from $N_{f}=2+1$ simulations~\cite{Bazavov:2019www}, and good agreement is observed. \\

\begin{table}[t]
\center
\small
\begin{tabular}{|c|c|c|c|c|c|c|c|c|c|c|c|}
\hline
\rule{0pt}{3ex}
 $\!\! N_{s}^{3}\times N_{\tau} \!\!$  & $\!\! a \ \text{[fm]} \!\!$ & $\!\! a \, m_{ud} \!\!$  & $\!\! T \ \text{[MeV]} \!\! $ &  $\!\! \text{Fit range} \!\!$ & $\!\!\!\! A \!\!\!\!$ & $\!\! m_{\pi}^{\text{scr}} \ \text{[MeV]} \!\!$ &  $\!\!\!\! B \!\!\!\!$ &  $\!\! m_{\pi^{*}}^{\text{scr}} \ \text{[MeV]} \!\!$ & $\!\! \chi^{2}/\text{d.o.f.} \!\! $   \\[0.5ex]
\hline
\rule{0pt}{3ex}
$32^3\times 12$	    & \!\! $0.075$ \!\! & \!\! $0.001$ \!\! & $220$ & \!\! $[2x_{3}^{*},L/2]$ \!\! & \!\!\!\! 0.114(7)  \!\!\!\!  & 776(17)  & \!\!\!\! 2.24(17) \!\!\!\!  & 3361(115) & 0.581 \\
$32^3\times 8$	    & \!\! $0.075$ \!\! & \!\! $0.001$ \!\! & $320$ & \!\! $[2x_{3}^{*},L/2]$ \!\! & \!\!\!\! 0.347(7)  \!\!\!\!  & 1781(5)  & \!\!\!\! 3.45(14) \!\!\!\!  & 4230(62)  & 0.686\\
$32^3\times 8$      & \!\! $0.065$ \!\! & \!\! $0.001$ \!\! & $380$ & \!\! $[2x_{3}^{*},L/2]$ \!\! & \!\!\!\! 0.325(9)  \!\!\!\!  & 2153(9)  & \!\!\!\! 3.49(15) \!\!\!\!  & 4998(78)  & 0.444 \\
$32^3\times 8$      & \!\! $0.051$ \!\! & \!\! $0.001$ \!\! & $480$ & \!\! $[2x_{3}^{*},L/2]$ \!\! & \!\!\!\! 0.343(11) \!\!\!\!  & 2860(11) & \!\!\!\! 3.61(16) \!\!\!\!  & 6430(108) & 0.314\\
$32^3\times 4$      & \!\! $0.075$ \!\! & \!\! $0.001$ \!\! & $660$ & \!\! $[x_{3}^{*},L/2]$  \!\! & \!\!\!\! 1.65(2)   \!\!\!\!  & 3816(4)  & \!\!\!\! 5.96(6)  \!\!\!\!  & 6012(39)  & 0.353\\
$32^3\times 4$      & \!\! $0.051$ \!\! & \!\! $0.001$ \!\! & $960$ & \!\! $[x_{3}^{*},L/2]$  \!\! & \!\!\!\! 1.52(4)   \!\!\!\!  & 5559(10) & \!\!\!\! 6.04(8)  \!\!\!\!  & 8604(77)  & 1.243\\[0.5ex]
\hline
\end{tabular}
\caption{Lattice parameters for the spatial pseudo-scalar correlator data from Ref.~\cite{Rohrhofer:2019qwq}, together with the best-fit parameters pertaining to the dashed lines in Fig.~\ref{z_corr_plot}.}
\label{tab:ensembles}
\end{table}
 
\noindent
Since the QCD pion in vacuum is a stable massive scalar particle, and corresponds to the lightest $T=0$ state, the considerations of Sec.~\ref{extract} apply, and one can use Eq.~\eqref{D_C_rel} to extract the damping factor associated with this state. Doing so, one obtains   
\begin{align}
D_{m_{\pi},\beta}(\vec{x})=\alpha_{\pi} \, e^{-\gamma_{\pi} |\vec{x}|},
\label{damp_m}
\end{align}
where $\alpha_{\pi}$ and $\gamma_{\pi}$ are $T$-dependent parameters. Note that a damping factor with this functional form was already observed in Ref.~\cite{Lowdon:2022ird} for pion states in the quark-meson model, in this case with correlators generated using a functional renormalisation group (FRG) approach\footnote{Recent analyses of pion spectral properties in the quark-meson model were also performed in Refs.~\cite{Tripolt:2013jra,Tripolt:2014wra,Helmboldt:2014iya}, based on an FRG analytic continuation approach.}. The parameter $\gamma_{\pi}$ was shown to control the thermal broadening of the pion states, and hence corresponds to the interaction energy between the pions and the thermal background.  For the sub-leading $\pi^{*}$ contribution it must also be the case that the corresponding damping factor has the form $D_{m_{\pi^{*}},\beta}(\vec{x})=\alpha_{\pi^{*}} \, e^{-\gamma_{\pi^{*}} |\vec{x}|}$. Therefore, the resolution of two independent exponential contributions to $\overline{C}_{\text{PS}}(x_{3})$ is consistent with the two lowest-lying particle states in the $T=0$ spectrum having exponential damping factors. Applying Eq.~\eqref{C_decomp_dom}, contributions of this form generate the general spatial correlator behaviour   
\begin{align}
C_{\text{PS}}(x_{3})=\sum_{i=\pi,\pi^{*}} \! C_{i}(x_{3}), \qquad  C_{i}(x_{3}) = \frac{\alpha_{i}}{2(m_{i}+\gamma_{i})}  e^{-(m_{i}+\gamma_{i})|x_{3}|}, 
\label{C_i}
\end{align} 
which implies the following relation between the screening masses, vacuum masses, and damping factor exponents for each particle contribution 
\begin{align}
m_{i}^{\text{scr}}(T) = m_{i} + \gamma_{i}(T), \quad\quad i=\pi, \, \pi^{*}. 
\label{m_screen}
\end{align}
Due to the restoration of Lorentz invariance in the $T\rightarrow 0$ limit, the parameters $\gamma_{i}$ must vanish, and hence
\begin{align}
C_{i}(x_{3}) \, \xlongrightarrow{T\rightarrow 0} \, \frac{\alpha_{i}(T\!=\!0)}{2m_{i}} e^{-m_{i}|x_{3}|},
\end{align}
as expected for the $T=0$ spatial correlator of a massive particle with weight $\alpha_{i}(T\!=\!0)$. Having determined the functional form of the $\pi$ and $\pi^{*}$ damping factors, one can now use the corresponding thermal spectral density 
\begin{align}
D_{\beta}(\vec{x},s)=  D_{m_{\pi},\beta}(\vec{x})\, \delta(s-m_{\pi}^{2}) + D_{m_{\pi^{*}},\beta}(\vec{x})\, \delta(s-m_{\pi^{*}}^{2}),
\label{D_PS}
\end{align}
together with the general representation in Eq.~\eqref{commutator_rep}, to establish how the $\pi$ and $\pi^{*}$ states contribute to the pseudo-scalar spectral function $\rho_{\text{PS}}(\omega,\vec{p})$. One finds that 
\begin{align}
\rho_{\text{PS}}(\omega,\vec{p})&= \epsilon(\omega) \left[ \theta(\omega^{2}-m_{\pi}^{2}) \,  \frac{4 \, \alpha_{\pi} \gamma_{\pi}  \sqrt{\omega^{2}-m_{\pi}^{2}}}{(|\vec{p}|^{2}+m_{\pi}^{2}-\omega^{2})^{2} + 2(|\vec{p}|^{2}-m_{\pi}^{2}+\omega^{2})\gamma_{\pi}^{2}+\gamma_{\pi}^{4} }            \right. \nonumber \\
& \quad\quad\quad\quad\quad\quad + \left. \theta(\omega^{2}-m_{\pi^{*}}^{2}) \,  \frac{4 \, \alpha_{\pi^{*}} \gamma_{\pi^{*}}  \sqrt{\omega^{2}-m_{\pi^{*}}^{2}}}{(|\vec{p}|^{2}+m_{\pi^{*}}^{2}-\omega^{2})^{2} + 2(|\vec{p}|^{2}-m_{\pi^{*}}^{2}+\omega^{2})\gamma_{\pi^{*}}^{2}+\gamma_{\pi^{*}}^{4} }     \right],
\label{commutator_PS}
\end{align}
which is the two-state generalisation of the result obtained in Ref.~\cite{Lowdon:2022ird}. Equation~\eqref{commutator_PS} therefore represents the spectral function that best describes the lattice data in the range $[2x_{3}^{*},L/2]$. Substituting Eq.~\eqref{commutator_PS} into the spatial correlator expression in Eq.~\eqref{C_rho} one recovers the result of Eq.~\eqref{C_i}, as expected. \\

\noindent
To interpret the physics of this result, let us focus on the zero-momentum spectral function $\rho_{\text{PS}}(\omega):= \rho_{\text{PS}}(\omega,\vec{p}=0)$, which has the form 
\begin{align}                                                   
\rho_{\text{PS}}(\omega)=   \epsilon(\omega) \left[ \theta(\omega^{2}-m_{\pi}^{2}) \, \frac{4\, \alpha_{\pi} \,  \gamma_{\pi} \sqrt{\omega^{2}-m_{\pi}^{2}}}{(\omega^{2}-m_{\pi}^{2}+\gamma_{\pi}^{2})^{2}} +\theta(\omega^{2}-m_{\pi^{*}}^{2}) \, \frac{4\, \alpha_{\pi^{*}} \,  \gamma_{\pi^{*}} \sqrt{\omega^{2}-m_{\pi^{*}}^{2}}}{(\omega^{2}-m_{\pi^{*}}^{2}+\gamma_{\pi^{*}}^{2})^{2}}   \right].
\label{spectral_pzero}
\end{align}
Particularly distinctive characteristics of Eq.~\eqref{spectral_pzero} are that the $\pi$ and $\pi^{*}$ contributions have sharp thresholds at their respective $T=0$ masses, and each particle contribution has a single peaked behaviour, with the peak maxima located at 
\begin{align}
\omega_{i}^{\text{peak}} = \sqrt{m_{i}^{2} + \tfrac{1}{3}\gamma_{i}^{2}}, \quad\quad  i=\pi, \, \pi^{*}.
\label{peak}
\end{align}   
Mathematically, the thresholds are a direct consequence of the discrete $\delta(s-m_{i}^{2})$ components in the ansatz of Eq.~\eqref{decomp}. It reflects the physical situation that the thermal background state $|\Omega_\beta\rangle $ is itself composed of pions, and therefore one has to inject an energy of at least $m_{\pi}$ or $m_{\pi^{*}}$ in order to create a zero-momentum excitation with $\pi$ or $\pi^{*}$ quantum numbers, respectively. The shift of the peaks to energies larger than $m_{\pi}$ and $m_{\pi^{*}}$ corresponds to the fact that in addition one has to supply the interaction energy of the particles with the thermal background in order to excite those states with maximal probability. Hence, the thresholds have a microscopic origin in the mass gaps of the constituents of the thermal system, whilst the peak positions are a collective feature of the macroscopic system. \\

\begin{figure}[t]
\begin{subfigure}{0.49\linewidth}
\centering
\includegraphics[width=1\columnwidth]{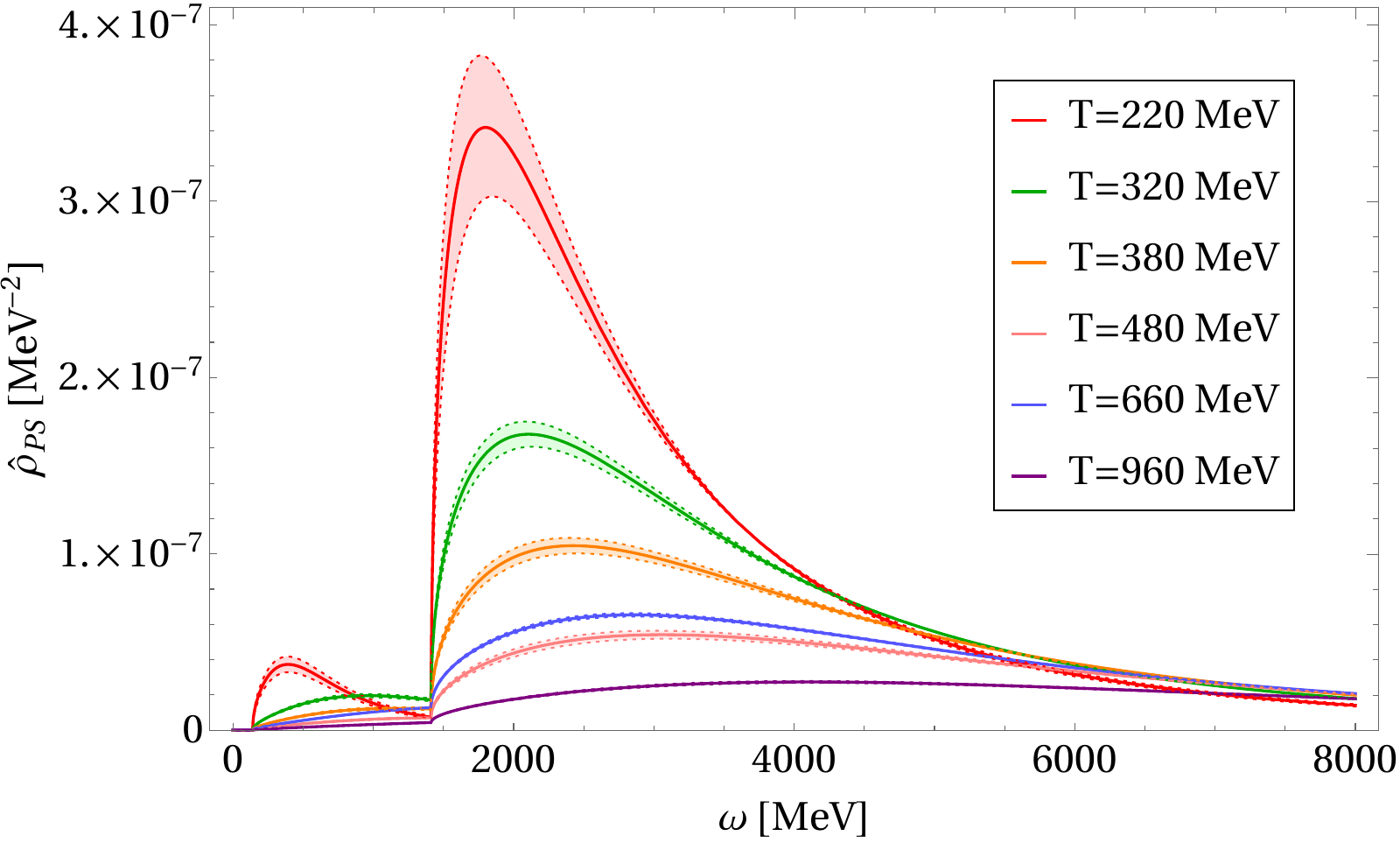}
\end{subfigure} 
\begin{subfigure}{0.495\linewidth}
\centering
\includegraphics[width=1\columnwidth]{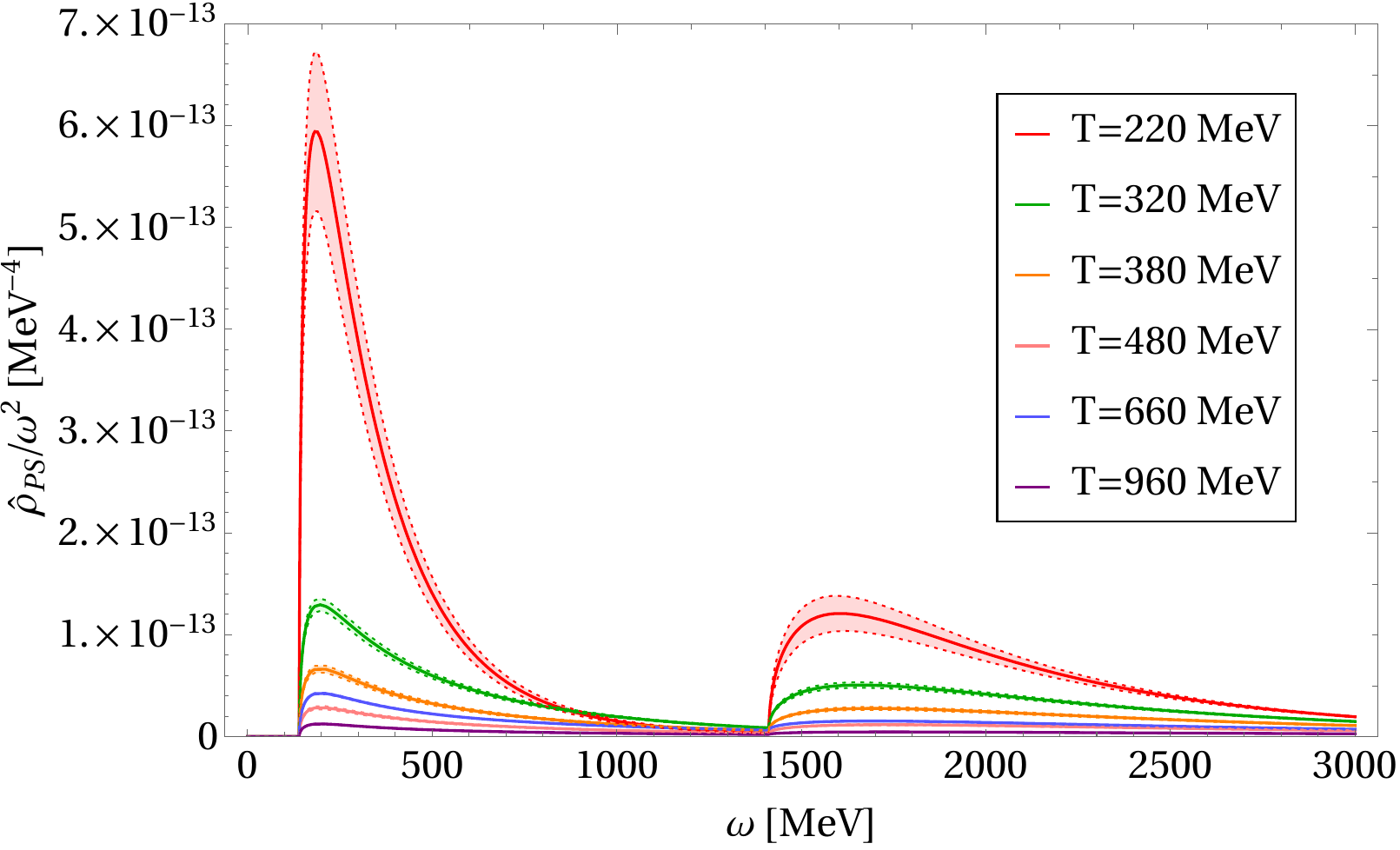}
\end{subfigure}
\caption{The unit-normalised pseudo-scalar spectral function $\hat{\rho}_{\text{PS}}(\omega)$ (left), and rescaled expression $\hat{\rho}_{\text{PS}}(\omega)/\omega^{2}$ (right), at different temperatures for $m_{\pi}=140$~MeV. The coloured bands indicate the 1$\sigma$ uncertainties.}
\label{spectral_plot} 
\end{figure}

\noindent
To complete our knowledge of the pseudo-scalar spectral function, we still need values for the coefficient parameters $\alpha_{\pi}$ and $\alpha_{\pi^{*}}$. In principle, these are fixed by the absolute normalisation of the correlation functions. However, this would require renormalised correlators, which are not available at present. Fortunately, by taking the ratio of the fitted coefficients this normalisation cancels, and one obtains the following expression for the relative weight of the $\pi$ and $\pi^{*}$ coefficients
\begin{align} 
\frac{\alpha_{\pi}}{\alpha_{\pi^{*}}}= \frac{m_{\pi}^{\text{scr}}  A}{m_{\pi^{*}}^{\text{scr}}  B  }. 
\label{fit_rel}
\end{align}
Since the spectral function has the functional form in Eq.~\eqref{spectral_pzero}, this implies that it satisfies the integral relation
\begin{align}
\int_{0}^{\infty} \frac{d\omega}{\pi} \, \omega \, \rho_{\text{PS}}(\omega) = \alpha_{\pi}+\alpha_{\pi^{*}}.
\label{int_rel}
\end{align}
By choosing to normalise the spectral function to one, it follows from Eqs.~\eqref{fit_rel} and~\eqref{int_rel} that the coefficients in the normalised spectral function $\hat{\rho}_{\text{PS}}(\omega)$ \textit{can} be calculated directly from the fitting parameters as
\begin{align}
\hat{\alpha}_{\pi} &= \frac{\alpha_{\pi}}{\alpha_{\pi}+\alpha_{\pi^{*}}}= \frac{A \, m_{\pi}^{\text{scr}}}{A \, m_{\pi}^{\text{scr}}+B \,m_{\pi^{*}}^{\text{scr}}}, \\
\hat{\alpha}_{\pi^{*}} &= \frac{ \alpha_{\pi^{*}}}{\alpha_{\pi}+\alpha_{\pi^{*}}} = \frac{B \, m_{\pi^{*}}^{\text{scr}}}{A \, m_{\pi}^{\text{scr}}+B \, m_{\pi^{*}}^{\text{scr}}}\;.
\end{align}
Using these relations, together with the fit parameters and Eq.~\eqref{spectral_pzero}, in the left plot of Fig.~\ref{spectral_plot} we evaluate $\hat{\rho}_{\text{PS}}(\omega)$ with its respective 1$\sigma$ error band for $m_{\pi}=140$~MeV at different temperatures. In the right plot of Fig.~\ref{spectral_plot} we also show $\hat{\rho}_{\text{PS}}(\omega)/\omega^{2}$, which is often studied in the literature. One sees that the rescaling by $\omega^{2}$ causes a separation of the $\pi$ and $\pi^{*}$ peaks, and their relative heights to change order.

\subsection{Systematics}

Before discussing the physics implications of our results, it is useful to assess the systematic uncertainties involved in their extraction. Starting with the lattice data we analysed here, it is worth repeating that domain wall fermions have good chiral properties at finite lattice spacing, and there are no doublers involved. This is important since the data we consider are not extrapolated to the continuum. The range of lattice spacings considered here (cf. Table~\ref{tab:ensembles}) is comparable to those used in a recent continuum extrapolated calculation of screening masses~\cite{Bazavov:2019www}, so finite lattice spacing effects are expected to be small. Similarly, for the spatial correlators, the box size in units of the largest correlation length (cf. Table~\ref{tab:ensembles}) is $m_{\pi}^{\text{src}} L \gtrsim 12$, and based on experience, finite-size effects in the correlators are exponentially small. \\

\noindent
A larger quantitative effect is caused by the fact that the bare quark mass is held fixed at $a m_{ud}=0.001$ in lattice units. Since there are different lattice spacings involved at different temperatures (cf. Table~\ref{tab:ensembles}), the system is not strictly staying on a so-called line of constant physics. Using the relation $m_{\pi}^{2}\sim m_{q}$ from chiral perturbation theory this amounts to a roughly 50\% uncertainty on the vacuum pion mass, and hence we set $m_{\pi}=140 \pm 70$~MeV. By using a linear extrapolation based on the $T=0$ excited pion lattice calculations of Ref.~\cite{Mastropas:2014fsa} we were able to estimate how this lattice uncertainty propagates to the mass of the first excited state, finding that $m_{\pi^{*}}=1412 \pm 38$~MeV. In order to establish the effect that these uncertainties have on the spectral function we varied the masses within their uncertainty ranges. Whilst we found that this produced some quantitative modifications of the peak heights on the order of 20\%, the overall qualitative characteristics of $\hat{\rho}_{\text{PS}}(\omega)$ were not significantly altered. Since the analysed lattice data correspond to $N_{f}=2$ QCD, one would also expect the spectral function to be quantitatively modified to some extent once strange quarks are included. \\

\noindent
Besides discrete particle components, the spectral function also contains continuous contributions, represented by $D_{c, \beta}(\vec{x},s)$ in Eq.~\eqref{D_decomp}. However, since the spatial correlator is directly related to the momentum-dependent spectral function via Eq.~\eqref{C_rho}, if these components did provide a significant contribution, particularly at small energies, they should be detectable in the fits. Together with the controllable effects discussed here, the largest systematic uncertainty in the present approach is the decomposition ansatz of Eq.~\eqref{decomp} proposed in Ref.~\cite{Bros:1992ey}, which encodes the significant influence of the vacuum states on the analytic structure of the spectral function. This ansatz is motivated by various physical considerations, as outlined in Secs.~\ref{intro} and~\ref{prelim}, and a general investigation will be the subject of future work. In the context of this study there exists a highly non-trivial test of the consistency of the extracted spectral functions, which is independent of how they are obtained. This will be discussed in Sec.~\ref{test}.

\subsection{A non-perturbative test of the spectral function}
\label{test}

It is apparent from Eqs.~\eqref{eq:corr} and~\eqref{C_rho} that the spectral function governs the behaviour of both spatial and temporal correlation functions. This therefore provides a straightforward and non-perturbative test of the consistency of any extracted spectral function: the spectral function determined from \textit{spatial} correlators must predict the corresponding \textit{temporal} ones, and vice versa. In our case, Ref.~\cite{Rohrhofer:2019qal} provides lattice data of the temporal correlator $C_{\text{PS}}(\tau):=\widetilde{C}_{\text{PS}}(\tau,\vec{p}=0)$ at $T=220$~MeV with the same lattice action and parameters as the spatial correlator data analysed here. From Eqs.~\eqref{eq:corr} and~\eqref{commutator_rep} it follows that
\begin{align}
C_{\text{PS}}(\tau) &= \int_{0}^{\infty}  \! \frac{d\omega}{2\pi} \,  \frac{\cosh\left[\left(\tfrac{\beta}{2}-|\tau| \right)\omega\right] }{\sinh\left(\tfrac{\beta}{2}\omega\right)} \, \rho_{\text{PS}}(\omega) \label{C_T_def} \\
&= \int_{0}^{\infty} \! ds \int_{0}^{\infty} \! \frac{d|\vec{u}|}{(2\pi)^{2}} \   \frac{|\vec{u}|^{2}\cosh\left[\left(\tfrac{\beta}{2}-|\tau| \right)\sqrt{|\vec{u}|^{2} +s}\right] }{\sqrt{|\vec{u}|^{2} +s} \, \sinh\left(\tfrac{\beta}{2}\sqrt{|\vec{u}|^{2} +s}\right)} \widetilde{D}_{\beta}(\vec{u},s), \label{PS_T_def}
\end{align}
and hence the extracted thermal spectral density in Eq.~\eqref{D_PS}, from which our $\rho_{\text{PS}}$ is calculated, can be used to predict the form of $C_{\text{PS}}(\tau)$, and compared with the data of Ref.~\cite{Rohrhofer:2019qal}. \\  

\begin{figure}[t] 
\centering
\includegraphics[width=0.6\columnwidth]{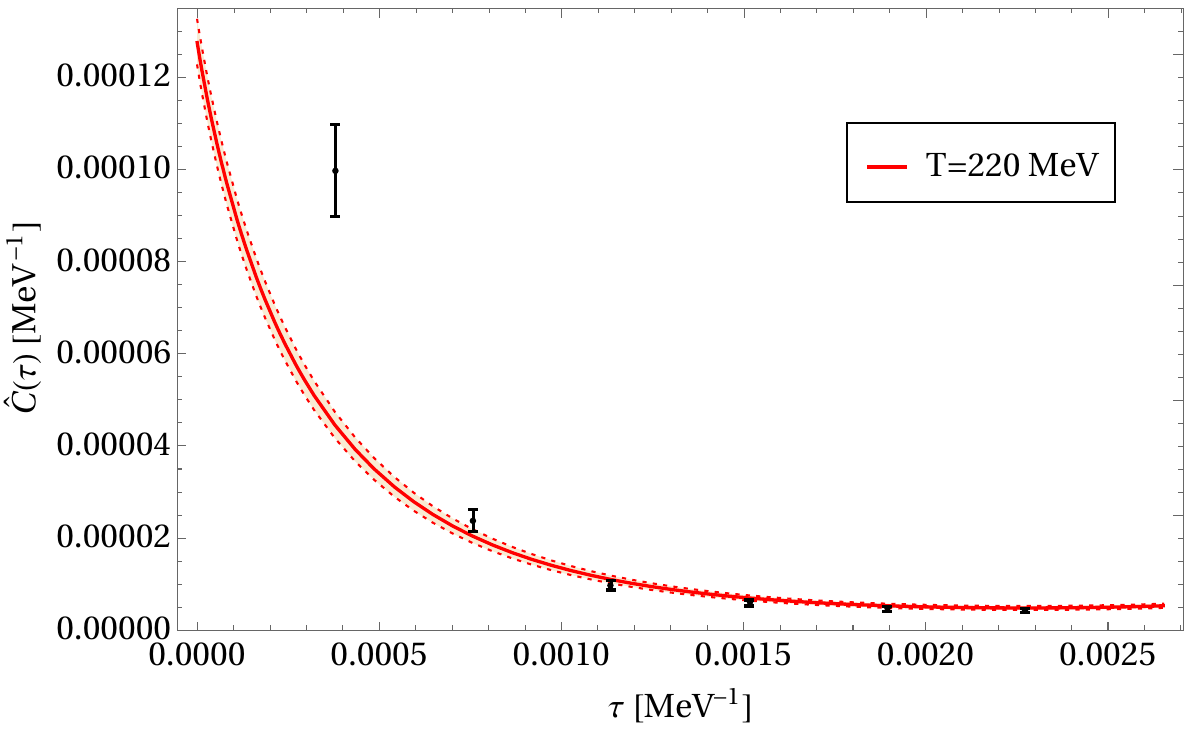}
\caption{The $\hat{C}_{\text{PS}}(\tau)$ prediction against the rescaled lattice data (black points) for $T=220$~MeV. The error bars and coloured band indicate the estimated combined uncertainties.} 
\label{T_corr_plot} 
\end{figure}

\noindent 
The lattice temporal correlator data from Ref.~\cite{Rohrhofer:2019qal} are normalised as $C_{\text{PS}}(\tau=\tau^{*})=1$, where $\tau^{*}=1/(T N_{\tau})$ is lattice separation one. In order to perform a comparison we need to ensure that both the spatial and temporal lattice correlators are normalised such that the zero-momentum spectral function has unit normalisation, as in Fig.~\ref{spectral_plot}. To do so, consider the derivative of $C_{\text{PS}}(\tau)$ for $\tau \rightarrow 0^{+}$, which due to Eq.~\eqref{C_T_def} gives
\begin{align}
\frac{d C_{\text{PS}}}{d\tau}(\tau\rightarrow 0^{+}) = -\int_{0}^{\infty}  \! \frac{d\omega}{2\pi} \omega \, \rho_{\text{PS}}(\omega).
\end{align}
Rescaling the temporal lattice data by $-2\frac{d C_{\text{PS}}}{d\tau}(\tau\rightarrow 0^{+})$ therefore ensures the proper normalisation, and we denote the corresponding correlator by $\hat{C}_{\text{PS}}(\tau)$. A final obstacle is the estimation of the derivative itself, since the data does not extend down to $\tau=0$. For this purpose we performed interpolations of the data based on Hermite polynomials and cubic splines, which were then extrapolated to $\tau=0$. Using the spread in results across the different interpolation methods we estimated the error on the derivative to be of the order of 10\%. In Fig.~\ref{T_corr_plot} we plot the $\hat{C}_{\text{PS}}(\tau)$ prediction from the spatial correlator data against the appropriately rescaled temporal lattice data. \\

\noindent
We see that the $\hat{C}_{\text{PS}}(\tau)$ prediction matches well with the rescaled lattice data. As $\tau$ becomes small the prediction begins to underestimate the lattice points. This is entirely expected, since $\hat{\rho}_{\text{PS}}(\omega)$ only contains information about the two lowest-lying states $\pi$ and $\pi^{*}$, whereas for small values of $\tau$ the higher-excited states such as the $\pi^{**}$ exert a greater influence over the behaviour of $\hat{C}_{\text{PS}}(\tau)$. Indeed, we observe quantitative agreement for $\tau \gtrsim 0.0007 \, \text{MeV}^{-1} \approx m_{\pi^{*}}^{-1}$, which is the smallest screening length resolved by the spatial correlator. Overall, the analysis in this section provides robust evidence that the spectral function in Eq.~\eqref{commutator_PS} is fully consistent with the available state-of-the-art lattice correlator data from Refs.~\cite{Rohrhofer:2019qwq,Rohrhofer:2019qal} on length scales $\ell \gtrsim m_{\pi^{*}}^{-1}$ at $T=220$ MeV.

\subsection{Dynamical implications}
\label{physics_res}

Apart from the appearance of thresholds related to the vacuum spectrum, which we discussed already, the most prominent feature in Fig.~\ref{spectral_plot} (left) is that the pion component of the spectral function gives a pronounced peak at the lowest temperature $T=220$~MeV, which corresponds to approximately $1.2 \, T_{\text{pc}}$. As the temperature increases, this peak eventually disappears due to two combined effects. Firstly, there is a broadening of the peak, as expected from the increasing collision rates in the medium. Secondly, the peak location, defined in Eq.~\eqref{peak}, begins to exceed the $\pi^{*}$ threshold at $\omega=m_{\pi^{*}}$, causing the pion peak to effectively disappear. The $\pi^{*}$ features a distinct peak up to around $T=660$~MeV, although the inclusion of higher-excited pion states, which are not captured by the lattice data, would modify this picture. Nevertheless, because the contributions of the higher-excited states have increasingly larger thresholds, their behaviour would not affect the pion contribution to $\hat{\rho}_{\text{PS}}(\omega)$. The emerging physical picture is that of sequential melting of the pion and its excitations with increasing temperature, and that this process happens gradually and at temperatures significantly larger than the chiral symmetry restoration scale $T_{\text{pc}}$. \\

\noindent
Note that the general characteristics of Fig.~\ref{spectral_plot} (right) are similar to those seen in some earlier reconstruction studies~\cite{Hatsuda:2000kb,Asakawa:2002xj}, where instead the spectral function was extracted from data of temporal correlators $C_{\Gamma}(\tau)$ via Eq.~\eqref{eq:corr}. In particular, these studies also see separated peaks of decreasing size for larger values of $\omega$, even when $T>T_{\text{pc}}$. A contrast with Fig.~\ref{spectral_plot} though is the absence of discrete thresholds, which is most likely because the reconstructed spectral functions are constrained to be smooth. Further meaningful comparisons can also be made with non-perturbative studies in the $\mathrm{O}(4)$ model~\cite{Engels:2009tv,Engels:2014bra,Florio:2021jlx}, and analyses of the pseudo-scalar spectral function in low-temperature QCD~\cite{Son:2001ff,Son:2002ci,Brandt:2014qqa,Brandt:2015sxa}. In the $\mathrm{O}(4)$ model the spectral functions also appear to possess peaks above the transition temperature, as in Fig.~\ref{spectral_plot}, and the peak locations are related to the screening masses~\cite{Engels:2009tv}. In QCD analyses~\cite{Son:2001ff,Son:2002ci,Brandt:2014qqa,Brandt:2015sxa} the pion screening mass $m_{\pi}^{\text{scr}}$ is similarly found to increase with temperature, and the notion of a quasi-particle mass is introduced in the low-temperature regime, motivated by the observation that the spatial correlator has a purely exponential decay at large distances. This exponential behaviour is explained by the presence of a discrete delta component in the spectral function whose argument is proportional to $m_{\pi}^{\text{scr}}$ at $\vec{p}=0$. However, in our case we demonstrate that an exponential spatial correlator can \textit{also} be achieved with exponential damping factors (cf. Eq.~\eqref{C_i}), and a resulting spectral function (Eq.~\eqref{commutator_PS}) with a quite different analytic structure.  \\

\noindent
Finally, we note that the picture of pions melting at temperatures significantly above the effective chiral symmetry restroration scale is fully consistent with the expectations based on chiral spin symmetry, which was observed to be approximately realised in a temperature range $T_{\text{pc} }  \lesssim T  \lesssim 3 T_{\text{pc}}$ on the same lattice configurations~\cite{Rohrhofer:2019qwq,Rohrhofer:2019qal}. As explained in those studies, chiral spin symmetry can only emerge dynamically in a regime where the chromoelectric quark gluon interactions dominate the quantum effective action of QCD. Refs.~\cite{Rohrhofer:2019qwq,Rohrhofer:2019qal} therefore concluded that in this intermediate temperature regime the thermal medium consists of hadron-like states where chiral symmetry is nearly restored, and quarks are still effectively bound. The spectral function of Fig.~\ref{spectral_plot} fully supports this picture. In Ref.~\cite{Glozman:2022lda} it was discussed how this chiral spin symmetric band smoothly extends to finite baryon density, where it may connect to quarkyonic matter in the cold and dense regime~\cite{McLerran:2007qj,Philipsen:2019qqm}, which is expected to contain chirally symmetric (parity-doubled) baryons. In this context, it would be most interesting to extend the present study to lattice correlators evaluated at imaginary baryon chemical potential, which can be computed without a sign problem, and understand the influence of small baryon densities on the spectral functions.

\subsection{The scalar spectral function and chiral symmetry restoration}

In Ref.~\cite{Rohrhofer:2019qwq} the correlators for several other iso-vector meson operators are given for the temperatures listed in Table~\ref{tab:ensembles}, in particular the Lorentz scalar, which contains contributions from the $a_{0}$ meson. The physical $a_{0}$ has a vacuum mass of roughly $980$~MeV, but this value will be lower in $N_{f}=2$ QCD because the $a_{0}$ is expected to have a sizeable $s\bar{s}$ component~\cite{Workman:2022ynf}. Moreover, the $a_{0}$ is unstable under strong decays in vacuum, and the effect of this instability also needs to be taken into account. For these reasons, we postpone the analysis of the other iso-vector meson correlators to future work. Nevertheless, a few qualitative observations can already be made. Due to the effectively restored $\mathrm{U}(1)_{\text{A}}$ symmetry for $T\gtrsim 220$~MeV, the scalar and pseudo-scalar spatial correlators are degenerate within the available accuracy of the data~\cite{Rohrhofer:2019qwq}. One might therefore expect the corresponding spectral function $\rho_{\text{S}}$ to be identical to $\rho_{\text{PS}}$ in this temperature regime. This, however, would only be the case for a theory with exact chiral symmetry, whereas for physical QCD, some care is in order. \\

\noindent
Although scalar and pseudo-scalar screening masses are approximately degenerate for $T>T_{\text{pc}}$, Eq.~\eqref{m_screen} demonstrates that this can be achieved even if the damping factor exponents $\gamma_{i}(T)$, and hence the spectral functions, are different, due to the unequal vacuum masses. If $\rho_{\text{S}}$ also has the same qualitative properties as $\rho_{\text{PS}}$, this implies that the low-energy behaviour of these spectral functions also cannot be equal since the pion and $a_{0}$ components will have different thresholds. It is well known from reconstruction studies that enormous accuracy of the correlation functions is necessary to resolve details of the spectral function, due to its folding with the thermal kernel in Eq.~\eqref{eq:corr}. In other words, slightly different spectral functions may well result in approximately degenerate correlators. This leads to the following natural picture: starting with different vacuum properties of the pion and $a_{0}$, the spectral functions $\rho_{\text{PS}}$ and $\rho_{\text{S}}$ will begin to move towards one another for higher temperatures on account of the thermal modifications encoded in their respective damping factors. In the temperature region $T \gtrsim T_{\text{pc}}$ investigated here, one still expects some differences, in particular around the pion threshold region, but plausibly these differences will only integrate to a small deviation between the respective correlators. As the temperature is further increased, the vacuum properties of the particles will be overpowered by the thermal modifications, and the spectral functions will effectively merge in the limit $T\rightarrow\infty$, as expected. The non-perturbative approach presented in this work therefore offers a viable path to unveil the details of chiral symmetry restoration and deconfinement in QCD.

\section{Conclusions} 
\label{concl}

Understanding the non-perturbative structure of Euclidean correlation functions at finite temperature is essential for obtaining a correct physical interpretation of how particles behave in the presence of a thermal medium. In this work, we utilise the general spectral representation satisfied by thermal two-point functions to derive a corresponding representation for the spatial correlator $C(x_{3})$ of real scalar fields. We find that if there is a mass gap to a stable vacuum particle state comprising the bulk of the medium, and a further gap to the continuum threshold, the large-$x_{3}$ behaviour of the correlator is dominated by discrete particle contributions, as expected from basic physical arguments. In such a regime, the spectral function can be extracted directly, avoiding the well-known inverse problem. \\ 

\noindent
Using these analytic results, we analyse lattice QCD data for the light-quark pseudo-scalar meson operator in the temperature range $220-960 \, \text{MeV}$, and extract the form of the corresponding spectral function $\rho_{\text{PS}}(\omega,\vec{p})$. As a non-trivial test, we demonstrate that the extracted spectral function reproduces the corresponding temporal lattice correlator data for $T= 220 \, \text{MeV}$. We find that the pion $\pi$ and its first-excited state $\pi^{*}$ dominate the behaviour of $\rho_{\text{PS}}$, and that the $\pi$ is clearly distinguishable in a range of temperatures above the chiral pseudo-critical temperature $T_{\text{pc}}$. Similar qualitative features should also be realised for the $a_{0}$ meson and its excitations, due to effective chiral symmetry restoration. Explicit relations between the vacuum masses, screening masses, and spectral function parameters are derived, and insights into the distinction between decays due to intrinsic vacuum instability and thermal broadening are also discussed. These findings suggest that non-perturbative effects continue to play a significant role above $T_{\text{pc}}$ even for hadronic states composed of light quarks, and are consistent with the expectations based on the approximate realisation of chiral spin symmetry in this temperature regime. Although this study focused on the spectral properties of correlators involving pseudo-scalar meson operators, our approach can in principle be generalised to meson states with higher spin, other hadronic states, as well as to regimes with non-vanishing baryon density. This work represents a step towards the analytic characterisation of non-perturbative in-medium effects in QCD.

\section*{Acknowledgements}
P.~L. would like to thank Shirley Li for useful discussions and input. The work of P.~L. and O.~P.~is supported by the Deutsche Forschungsgemeinschaft (DFG, German Research Foundation) through the Collaborative Research Center CRC-TR 211 ``Strong-interaction matter under extreme conditions'' -- Project No. 315477589-TRR 211. O.~P.~also acknowledges support by the State of Hesse within the Research Cluster ELEMENTS (Project ID 500/10.006).

\bibliographystyle{JHEP}

\bibliography{refs}

\end{document}